\documentclass[notitlepage,aps,onecolumn,superscriptaddress,nofootinbib,floatfix,onecolumn]{revtex4-1}
\pdfoutput=1

\usepackage{dcolumn}
\usepackage{bm}

\usepackage[dvips]{graphicx}
\usepackage{float}
\usepackage{epsfig}
\usepackage{graphicx}
\usepackage{longtable}
\usepackage{rotating}
\usepackage{ulem}
\usepackage{amsmath,amssymb,color,hyperref}
\usepackage{txfonts}
\usepackage{mathtools}
\usepackage{tabularx}
\usepackage{booktabs}
\usepackage{xspace}

\def\ttt#1{\texttt{\small #1}}

\newcommand{\sqrts}{\sqrt{\rm s}}
\newcommand{\pp}{p-p}
\newcommand{\ppbar}{p-${\rm \bar p}$}
\newcommand{\epem}{e^+e^-}
\newcommand{\pT}{\rm p_{_{\rm T}}}
\newcommand{\dNdeta}{\rm dN_{\rm ch}/d\eta}
\newcommand{\dNdetaZero}{\rm dN_{\rm ch}/d\eta|_{\eta=0}}

\newcommand{\dNdetaZeroNSD}{\rm dN_{\rm ch}^{^{\rm NSD}}/d\eta|_{\eta=0}}
\newcommand{\dEdeta}{\rm dE/d\eta}
\newcommand{\dNdpT}{\rm dN_{\rm ch}/dp_{_{\rm T}}}
\newcommand{\PNch}{\rm P(N_{\rm ch})}
\newcommand{\meanpt}{\rm \left< p_{\rm T} \right>}
\newcommand{\Lambdaqcd}{\Lambda_{_{\rm QCD}}}
\newcommand{\sigmainel}{\sigma_{\rm inel}}
\newcommand{\sigmahard}{\sigma_{_{\rm pQCD}}}

\newcommand{\Pom}{\mathbb{P}}

\newcommand{\phojet}{\textsc{phojet}}
\newcommand{\epos}{\textsc{epos}}
\newcommand{\eposlhc}{\textsc{epos-lhc}}
\newcommand{\qgsjet}{\textsc{qgsjet}} 
\newcommand{\qgsjetII}{\textsc{qgsjet-ii}} 
\newcommand{\sibyll}{\textsc{sibyll}}

\newcommand{\pythia}{\textsc{pythia}}
\newcommand{\herwig}{\textsc{herwig}}
\newcommand{\lhapdf}{\textsc{lhapdf}}

\def\cO#1{{{\cal{O}}}\left(\rm #1\right)} 

\newcommand*{\cm}{c.m.\@\xspace}
\newcommand*{\ie}{i.e.\@\xspace}
\newcommand*{\eg}{e.g.\@\xspace}

\begin{document}

\title{Global properties of proton-proton collisions at $\sqrts$~=~100~TeV}
 
\author{David~d'Enterria}
\affiliation{EP Department, CERN, 1211 Geneva, Switzerland}
\author{Tanguy Pierog}
\affiliation{Institut f\"ur Kernphysik, Karlsruhe Institute of Technology, Postfach 3640, 76021 Karlsruhe, Germany}

\begin{abstract}
\noindent
The global properties of the final states produced in hadronic interactions of protons at centre-of-mass
energies of future hadron colliders (such as FCC-hh at CERN, and SppC in China),
are studied. The predictions of various Monte Carlo (MC) event generators used in collider physics (\pythia~6,
\pythia~8, and \phojet) and in ultrahigh-energy cosmic-rays studies (\epos, and \qgsjet) are compared. Despite
their different underlying modeling of hadronic interactions, their predictions for proton-proton (\pp)
collisions at $\sqrts$~=~100~TeV are quite similar. The average of all MC predictions (except \phojet) for
the different observables are: 
(i) \pp\ inelastic cross sections $\sigmainel$~=~105~$\pm$~2~mb;
(ii) total charged multiplicity $\rm N_{_{\rm ch}}$~=~150~$\pm$~20;
(iii) charged particle pseudorapidity density at midrapidity $\dNdetaZero = 9.6 \pm 0.2$;
(iv) energy density at midrapidity $\rm dE/d\eta|_{\eta=0} =  13.6 \pm 1.5$~GeV, and
$\rm dE/d\eta|_{\eta=5} = 670 \pm 70$~GeV at the edge of the central region; and
(v) average transverse momenta at midrapidities $\meanpt = 0.76 \pm 0.07$~GeV/c.
At midrapidity, 
\epos\ and \qgsjetII\ predict larger per-event multiplicity probabilities at very low ($\rm N_{\rm ch}<3$)
and very high ($\rm N_{\rm  ch}>100$) particle multiplicities, whereas \pythia~6 and 8 feature higher yields
in the intermediate region $\rm N_{\rm ch}\approx$~30--80. 
These results provide useful information for the estimation of the detector occupancies and energy
deposits from pileup collisions at the expected large FCC-hh/SppC luminosities.
\end{abstract}


\maketitle

\section{Introduction}

The Future Circular Collider (FCC) is a post-LHC project in a new 100-km tunnel under consideration at CERN, 
that would provide hadron and $\epem$ collisions at much higher energies and luminosities than studied so
far. Its key scientific goals are the complete exploration of the Higgs sector of the Standard 
Model (SM), and a significant extension in searches of physics beyond the SM via direct or indirect
measurements~\cite{Mangano16a,Mangano16b,Mangano16c}. The FCC-hh will deliver proton-proton (\pp) collisions at a centre-of-mass (\cm) energy of
$\sqrts$~=~100~TeV with integrated luminosities at the level of several 100~fb$^{-1}$ per year or
above~\cite{Benedikt:2015poa}. Ongoing studies exist on the detector requirements needed to carry out the
planned measurements under running conditions involving 
$\cO{200-1000}$ simultaneous \pp\ collisions per bunch crossing. Similar studies are under consideration 
in the context of the Super proton-proton Collider (SppC) promoted by IHEP in China\cite{CEPC-SPPCStudyGroup:2015csa}.
This work presents a study of the average properties of multiparticle production in \pp\ collisions at FCC-hh/SppC energies, 
of usefulness, among others, for the estimation of the expected occupancies and energy deposits in the planned FCC-hh/SppC detectors.\\

Inclusive particle production in high-energy hadronic collisions receives contributions from 
``soft'' and ``hard'' interactions, loosely separated by the virtuality 
of the underlying $t$-channel exchanges. Soft (hard) processes involve partons of virtualities $q^2$ typically
below (above) a scale $\rm Q_0^2\approx$~1--2~GeV. Semihard parton-parton scatterings around $\rm Q_0$,
dominate the inelastic hadron production cross sections for \cm\ energies above a few 
hundreds GeV, whereas soft scatterings dominate at lower energies ($\sqrts\lesssim$~20~GeV) where few hadrons
with low transverse momenta $\pT$ are produced. On the one hand, hard processes 
can be theoretically described within perturbative Quantum Chromodynamics (pQCD) in a collinear-factorized
approach through the convolution of parton distribution functions (PDFs) and matrix elements for the underlying
parton-parton collisions subprocesses. The scattered quarks and gluons produce then collimated bunches of final-state
hadrons (jets) through a parton branching process dominated by perturbative splittings described by the
Dokshitzer-Gribov-Lipatov-Altarelli-Parisi (DGLAP) equations~\cite{dglap1,dglap2,dglap3}, followed by non-perturbative
hadronization when the parton virtuality is below $\cO{1~GeV}$. On the other hand, soft processes have 
momenta exchanges not far from $\Lambdaqcd \approx$~0.2~GeV and, although they cannot be treated within pQCD, 
basic quantum field-theory principles  {\textemdash} such as unitarity and analyticity of scattering amplitudes as implemented in
Gribov's Reggeon Field Theory (RFT)~\cite{Gribov:1968fc} and exemplified \eg\ in the original Dual Parton
Model~\cite{Capella:1992yb} {\textemdash} give a decent account of their cross sections in terms of the exchange of
virtual quasi-particle states (Pomerons and Reggeons).
Given the extended composite nature of hadrons, even at asymptotically large energies, a non-negligible
fraction of inelastic \pp\ interactions involve soft ``peripheral'' scatterings.  
The Pomeron ($\Pom$) contribution, identified perturbatively with a colour-singlet multigluon exchange, 
dominates over those from secondary Reggeons (virtual mesons)
and is responsible for 
diffractive dissociation accounting for a noticeable fraction, about a fourth, of the  
total inelastic cross section at high energies~\cite{Donnachie:2002en,Khachatryan:2015gka}.\\

The general-purpose Monte Carlo (MC) models used in high-energy collider physics, such as
\pythia~6~\cite{Sjostrand:2006za}, \pythia~8~\cite{Sjostrand:2007gs}, \herwig++~\cite{Bahr:2008pv},
and \textsc{sherpa}~\cite{Gleisberg:2008ta}, are fully based on a pQCD framework which then incorporates soft
diffractive scatterings in a more or less ad hoc manner. In contrast, MC models commonly used in
cosmic-ray physics~\cite{d'Enterria:2011kw} such as \epos~\cite{Werner:2005jf,Pierog:2009zt,Pierog:2013ria}, 
\qgsjet~01~\cite{Kalmykov:1993qe,Kalmykov:1997te},
\qgsjetII~\cite{Ostapchenko:2005nj,Ostapchenko:2004ss,Ostapchenko:2007qb,Ostapchenko:2010vb} and
\sibyll~\cite{Ahn:2009wx}, as well as \phojet~\cite{Engel:1994vs,Engel:1995yda,Engel:1995sb}
mostly used for collider environments, are fully-based on the RFT approach. The latter MCs start off from a
construc\-tion of the hadron-hadron elastic scattering amplitude to determine the total, elastic and inelastic
(including diffractive) cross sections, 
extended to include hard processes via ``cut (hard) Pomerons'' (also known as ``parton ladder'') diagrams.\\ 

At increasingly larger \cm\ energies, the inelastic cross section receives major contributions from the region
of low parton fractional momenta ($x=p_{_{\rm parton}}/p_{_{\rm hadron}}$), where the
gluon distribution rises very fast. As a matter of fact, at $\sqrts$~=~100~TeV the partonic cross section
saturates the total inelastic cross section (\ie\ $\sigmahard\approx\sigmainel\approx$~100~mb) at 
momenta $\pT\approx$~10~GeV/c, 50 times larger than  $\Lambdaqcd$. Such a ``divergent'' behaviour (taking
place {\it well} above the infrared regime) is solved by reinterpreting this observation as a consequence of
the increasing number of multiparton 
interactions (MPI) occurring in a single \pp\ collision. Multiple scattering is naturally incorporated in the
RFT models through the ``eikonalization'' of multi-Pomeron exchanges that unitarize the cross sections, whereas
\pythia\ eikonalises multiparton exchanges, supplemented with an impact-parameter (Glauber-like) description of
the proton~\cite{Sjostrand:1987su}. 
The energy evolution of such MPI and low-$x$ effects is implemented phenomenologically in all MCs through a
transverse momentum cutoff $\rm Q_0$ of a few GeV that tames the fastly-rising $1/\pT^4$ minijet cross section
(\eg\ in \pythia\ the cutoff is introduced through a multiplicative $1/(\pT^2+Q_0^2)^2$ factor). This $\rm Q_0$
regulator is often defined so as to run with \cm\ energy following a slow power-law (or logarithmic)
dependence, closely mimicking the ``saturation scale'' $\rm Q_{\rm sat}$ that controls the onset of non-linear
(gluon fusion) effects saturating the growth of the PDFs as $x\to$~0~\cite{Gribov:1984tu}.
Last but not least, all MC generators, both based on pQCD or RFT alike, use parton-to-hadron fragmentation
approaches fitted to the experimental data  {\textemdash} such as  the Lund string~\cite{Andersson:1983ia}, area law~\cite{Artru:1974hr} or cluster 
hadronization~\cite{Marchesini:1991ch} models {\textemdash}  to hadronize the coloured degrees of freedom once their 
virtuality 
evolves below $\cO{1~GeV}$.\\

In this paper, we compare the basic properties of the so-called ``minimum bias'' (MB) observables
characterizing the final states produced in proton-proton collisions at $\sqrts$~=~100~TeV, predicted by pQCD-
and RFT-based hadronic interaction models. The MB term refers commonly to inelastic interactions
experimentally measured using a generic minimum-bias trigger that accepts a large fraction of the particle
production cross section by requiring a minimum activity in one or various detectors. In some cases we
present also results for the so-called ``non single-diffractive'' (NSD) events, mimicking the typical
experimental requirement of a two-arm trigger with particles in opposite hemispheres to eliminate
backgrounds from beam-gas collisions and cosmic-rays. Such NSD topology reduces significantly the
detection rate of (single) diffractive collisions characterized by the survival of one of the colliding
protons and particle production in just one hemisphere.
The phenomenological setup of our study is described in Section \ref{sec:th}, the predictions of the different
MCs for basic inclusive particle production observables  {\textemdash} such as the inelastic cross section $\sigmainel$, the
particle and energy densities as a function of pseudorapidity $\dNdeta$ and $\dEdeta$, the per-event multiplicity
distribution $\PNch$, and the transverse momentum distribution $\dNdpT$ (and associated mean transverse momenta
$\meanpt$) {\textemdash}  are presented in Section~\ref{sec:results}, and the main conclusions are summarized in
Section~\ref{sec:summary}.

\section{Theoretical setup}
\label{sec:th}


The basic ingredients of the \pythia~6 and 8 event generators are leading-order (LO) pQCD $2\to 2$ matrix elements, 
complemented with initial- and final-state parton radiation (ISR and FSR), folded with PDFs (interfaced here via
the \lhapdf\ v6.1.6 package~\cite{Bourilkov:2006cj}), and the Lund string model for parton hadronization. The
decomposition of the inelastic cross section into non-diffractive and diffractive components is based on a Regge model~\cite{Schuler:1993wr}. 
In this work we use the \pythia\ event generator in two flavours: the Fortran version
6.428~\cite{Sjostrand:2006za}, as well as the C++ version \pythia\ 8.17~\cite{Sjostrand:2007gs}.
We consider two different ``tunes'' of the parameters governing the non-perturbative and semihard dynamics:
ISR and FSR showering, MPI, beam-remnants, FS colour-reconnection, and hadronization. For \pythia\ 6.4 we 
use the Perugia-350 tune~\cite{Skands:2010ak}, whereas for \pythia\ 8 we use the Monash 2013 tune
(\ttt{Tune:ee=7; Tune:pp=14})~\cite{Skands:2014pea}. Both sets of parameters (Table~\ref{tab:pythiaTunes}) have been
obtained from recent (2011 and 2013 respectively) analysis of 
MB, underlying-event (UE), and/or Drell-Yan data in \pp\ collisions at $\sqrts$~=~7~TeV.

\begin{table}[htbp]
\centering
{\footnotesize
\begin{tabular}{lccccccccc}\hline
Version & Tuning          & Diffraction & \multicolumn{3}{c}{Semihard dynamics} & \multicolumn{2}{c}{Initial state} & \multicolumn{2}{c}{Final state}  \\ 
        & (\ttt{PYTUNES}) &             & $\sqrt{s_0}$ & $\rm Q_0$ & power $\epsilon$ &   PDF  & transv. overlap
        & \hspace{0.2cm} colour reconnection & hadronization \\ 
\hline
6.428 & Perugia 2011 (350)& Regge-based~\cite{Schuler:1993wr} & 7~TeV & 2.93 GeV & 0.265 & CTEQ5L       &
$\exp(-r^{1.7})$  & strong & Lund model fits (2011)\\ 
8.170 & Monash 2013 (14)  & improved~\cite{Rasmussen:2015iid} & 7~TeV & 2.28 GeV & 0.215 & NNPDF2.3 LO  &
$\exp(-r^{1.85})$  & strong & Lund model fits (2013)\\ 
\hline
\end{tabular}
}
\caption{Comparison of the various ingredients controlling the non-perturbative and semihard (MPI, parton saturation)
dynamics in the two \pythia\ MCs used in this work. See text for details.}
\label{tab:pythiaTunes}
\end{table}

For the initial-state, \pythia~6 (Perugia 350) uses the CTEQ5L parton densities~\cite{Lai:1999wy} and
\pythia~8 (Monash) the NNPDF2.3 LO set~\cite{Ball:2013hta}, whereas for the description of the transverse parton
density, both models use an exponential-of-power profile of the \pp\ overlap function, $\exp(-r^{n})$, with slightly
different exponents ($n$~=~1.7 and 1.85 respectively). The \pythia~6 choice results in a broader \pp\ overlap 
which thereby enhances the fluctuations in the number of MPI relative to the Monash-2013 choice.
The energy evolution of the MPI cutoff is driven by $\rm Q_0^2(s)=Q_0^2(s_0)\cdot(s/s_0)^\epsilon$, 
with the parameters quoted in Table~\ref{tab:pythiaTunes}.
Given that the generation of additional parton-parton interactions in the UE is suppressed below $\rm Q_0$, a 
{\it higher} scaling power $\epsilon$ implies a {\it slower} increase of the overall hadronic activity. Thus,
the Monash tune results in a slower evolution of $\rm Q_0$, yielding larger MPI activity at 100~TeV compared to 
the Perugia tune.
The treatment of diffraction has improved in \pythia~8 compared to 6. In the former, a diffractive system is
viewed as a Pomeron-proton collision, including hard scatterings subject to all the same ISR/FSR 
and MPI dynamics as for a ``normal'' parton-parton process~\cite{Rasmussen:2015iid}.
For the final-state, the two tunes have strong final-state colour reconnections (implemented through different
models~\cite{Skands:2007zg,Sjostrand:2014zea}), which act to reduce the number of final-state
particles (for a given $\rm Q_0$ value) or, equivalently, lower the $\rm Q_0$ value that is required to reach
a given average final-state multiplicity. 
The Lund hadronization parameters for light- and heavy-quarks have been updated in \pythia~8 
compared to \pythia~6 by refitting updated sets of LEP and SLD data~\cite{Skands:2014pea}.\\


The RFT-based models used in this work differ in various approximations for the collision configurations (\eg\
the distributions for the number of cut Pomerons, and for the energy-momentum partition among them), the
treatment of diffractive and semihard dynamics, the details of particle production from string fragmentation,
and the incorporation or not of other final-state effects (Table~\ref{tab:RFT_MCs}). 
Whereas the RFT approach is applied using only Pomerons and Reggeons in the case of \qgsjet\ and \phojet, 
\epos\ extends it to include partonic constituents~\cite{Drescher:2000ha}. In the latter case, this is done
with an exact implementation of energy sharing between the different constituents of a hadron at the amplitude
level. The evolution of the parton ladders from the projectile and the target side towards the centre (small
$x$) is governed by the DGLAP equations.
For the minijet production cutoff, \phojet\ uses dependence of the form $\rm Q_0(s) \sim Q_0 + C \cdot
\log(\sqrts)$, whereas \epos\ and \qgsjetII\ use a fixed value of $\rm Q_0$. The latter MC resums dynamically
low-$x$ effects through enhanced diagrams corresponding to multi-Pomeron
interactions~\cite{Ostapchenko:2005nj,Ostapchenko:2006vr,Ostapchenko:2006nh}. 
In that framework, high mass diffraction and parton saturation are
related to each other, being governed by the chosen multi-Pomeron vertices,
leading to impact-parameter and density-dependent saturation at low momenta~\cite{Ostapchenko:2005yj}.
LHC data were used to tune the latest \qgsjetII-04 release~\cite{Ostapchenko:2010vb} shown here. 
\epos\, on the other hand, uses the wealth of RHIC proton-proton and nucleus-nucleus 
data to parametrize the low-$x$ behaviour of the parton densities in a more phenomenological way~\cite{Werner:2005jf}
(correcting the $\Pom$ amplitude used for both cross section and particle production).
The \epos\ MC is run with the LHC tune~\cite{Pierog:2013ria} which includes collective final-state string
interactions which result in an extra radial flow of the final hadrons produced in more central \pp\
collisions. Among all the MC models presented here, \phojet\ is the only one which does not take into account
any retuning using LHC data (its last parameter update dates from year 2000).

\begin{table}[htbp]
\centering
\begin{tabular}{lcccc}\hline
Model (version)                        & Diffraction  & \multicolumn{2}{c}{Semihard dynamics}   & Final state \\
                                       &              &    $\rm Q_0$      &  Evolution        &       \\\hline
\eposlhc~\cite{Pierog:2013ria}         & effective diffractive $\Pom$  & 2.0 GeV  & power-law corr. of $\Pom$ & collective flow + area law hadronization  \\
\qgsjetII-04~\cite{Ostapchenko:2005nj,Ostapchenko:2004ss,Ostapchenko:2007qb} & $\Pom$ cut-enhanced graphs + G.-W.~\cite{Good:1960ba}& 1.6 GeV  & enhanced $\Pom$-graphs  & simplified string hadronization\\
\phojet\ 1.12~\cite{Engel:1994vs,Engel:1995yda} & G.-W. model~\cite{Good:1960ba} & 2.5 GeV & $\rm Q_0(s)\propto\log(\sqrts)$ & hadronization via \pythia\ 6.115 \\ \hline
\end{tabular}
\caption{Comparison of the main ingredients controlling the non-perturbative and semihard dynamics
present in the RFT-based event generators used in this work.}
\label{tab:RFT_MCs}
\end{table}

The results are presented, in the case of \pythia~6 and 8, for primary charged particles, defined as all
charged particles produced in the collision including the products of strong and electromagnetic decays but
excluding products of weak decays, obtained by decaying all unstable particles\footnote{\pythia\ 6.4:
  \ttt{MSTJ(22)=2,PARJ(71)=10}. \pythia\ 8: \ttt{ParticleDecays:limitTau0 = on}, \ttt{ParticleDecays:tau0Max = 10}.} 
for which $c\tau<$~10~mm. For the RFT MCs, unless stated otherwise, the results correspond to the primary 
charged hadrons (with the same $c\tau$ requirement) but without charged leptons which, nonetheless,
represent a very small correction (amounting to about 1.5\% of the total charged yield, mostly from the Dalitz
$\pi^0$ decay). Unless explicitly stated, no requirement on the minimum $\pT$ of the particles is applied in
any of the results presented.

\section{Results}
\label{sec:results}

\subsection{Inelastic \pp\ cross section}

\begin{figure}[htpb!]
\centering
\includegraphics[width=0.5\textwidth]{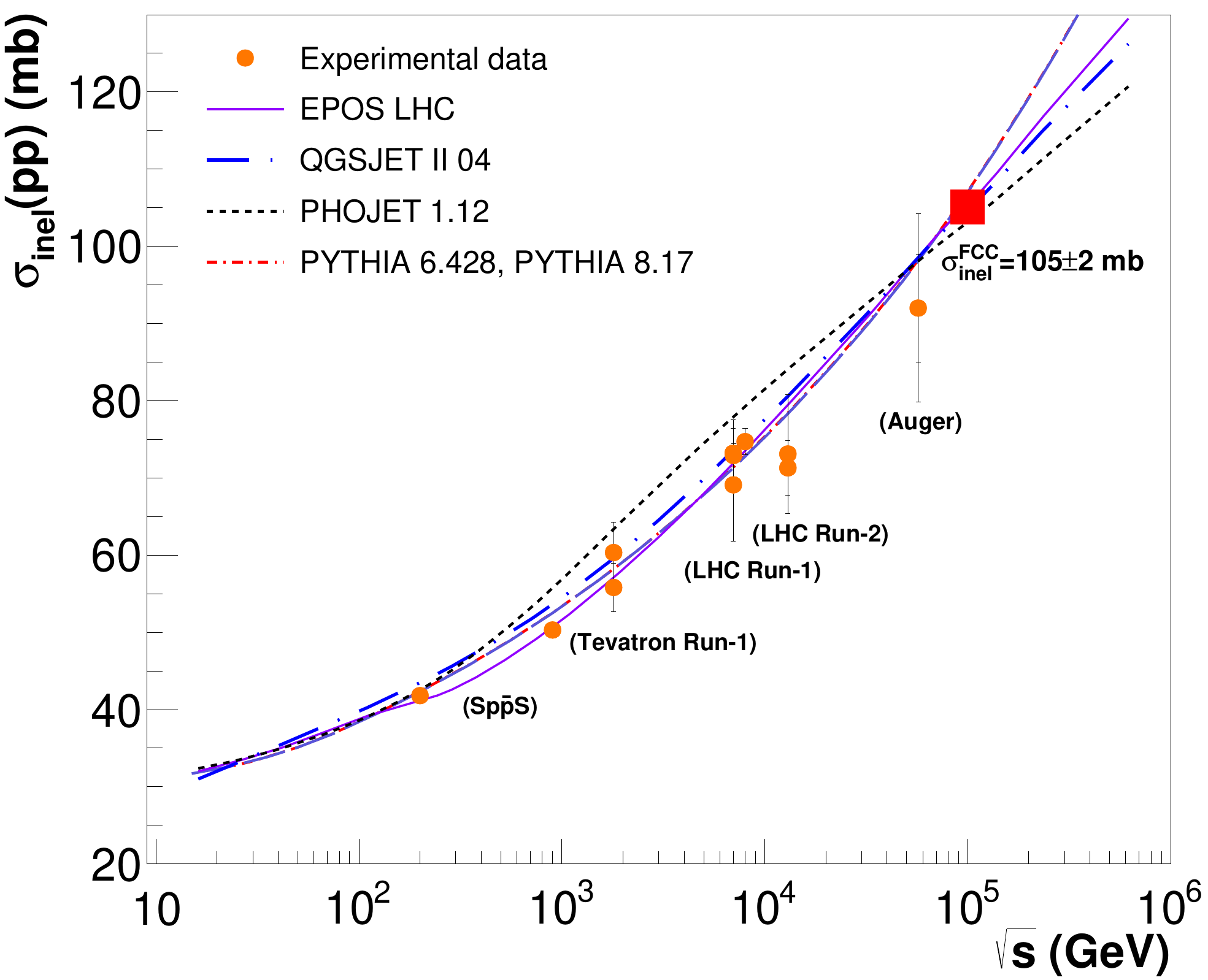}
\caption{Inelastic p-p cross section $\sigmainel$ as a function of \cm\ energy in the range
  $\sqrts~\approx$~10~GeV--500~TeV. Experimental data points at various collider and cosmic-ray
  energies~\cite{Alner:1986iy,Amos:1991bp,Abe:1993xy,Abelev:2012sea,Aad:2011eu,Antchev:2011vs,Antchev:2013iaa,Antchev:2013paa,atlas13TeV,Chatrchyan:2012nj,cms13TeV,Auger:2012wt} 
  are compared to the predictions of \eposlhc, \qgsjetII-04, \phojet~1.12, and \pythia~(both 6.428 and 8.17
  predict the same dependence). The red box indicates the average prediction of all models at 100~TeV.}
\label{fig:sigma_pp_vs_sqrts}
\end{figure}

The most inclusive quantity measurable in \pp\ collisions is the total hadronic cross section 
$\sigma_{\rm tot}$ and its separation into elastic and inelastic (and, in particular, diffractive) components. 
In both \pythia\ 6 and 8, the total hadronic cross section is calculated using the Donnachie-Landshoff
parametrisation~\cite{Donnachie:1992ny}, including Pomeron and Reggeon terms, whereas the elastic and
diffractive cross sections are calculated using the Schuler-Sj\"ostrand model~\cite{Schuler:1993wr}. 
The predictions for the inelastic cross sections in \pp\ at $\sqrts$~=~100~TeV, obtained simply from
$\sigma_{\rm tot} - \sigma_{\rm el}$, yield basically the same value, $\sigmainel \approx 107$~mb, for both
\pythia\ 6 and 8. The RFT-based MCs, based on $\Pom$ amplitudes, predict slightly lower values:
$\sigmainel = 105.4, 104.8, 103.1$~mb for \eposlhc, \qgsjetII\ and \phojet\ respectively. 
The $\sqrts$ dependence of the inelastic cross section predictions is shown in 
Fig.~\ref{fig:sigma_pp_vs_sqrts} together with the available data from \ppbar\ 
(UA5~\cite{Alner:1986iy}, E710~\cite{Amos:1991bp} and CDF~\cite{Abe:1993xy}) and \pp\
(ALICE~\cite{Abelev:2012sea}, ATLAS\cite{Aad:2011eu,atlas13TeV}, CMS~\cite{Chatrchyan:2012nj,cms13TeV},
TOTEM~\cite{Antchev:2011vs,Antchev:2013iaa,Antchev:2013paa}) colliders, as well as the AUGER result at $\sqrts$~=~57~TeV derived from
cosmic-ray data~\cite{Auger:2012wt}. Interestingly, all model curves cross at about $\sqrts\approx$~60~TeV,
and predict about the same inelastic cross section at the nominal FCC-hh/SppC \pp\ \cm\ energy of 100~TeV.
A simple average among all predictions yields $\sigmainel(\rm 100\;TeV) = 105.1 \pm 2.0$~mb, whereas larger
differences in the energy evolution of $\sigmainel$ appear above the $\sqrts\approx$~300~TeV, \ie\ around and
above the maximum energy observed so far in high-energy cosmic rays impinging on Earth
atmosphere~\cite{d'Enterria:2011kw}. 
The expected increase in the inelastic \pp\ cross section at 100~TeV is about 45\% compared to the LHC
results at 13~TeV ($\sigmainel$~=~73.1~$\pm$~7.7~mb~\cite{atlas13TeV}, and (preliminary)
71.3~$\pm$~3.5~mb~\cite{cms13TeV}). 

\subsection{Particle pseudorapidity density}

\begin{figure}[htpb!]
\centering
\includegraphics[width=0.49\textwidth]{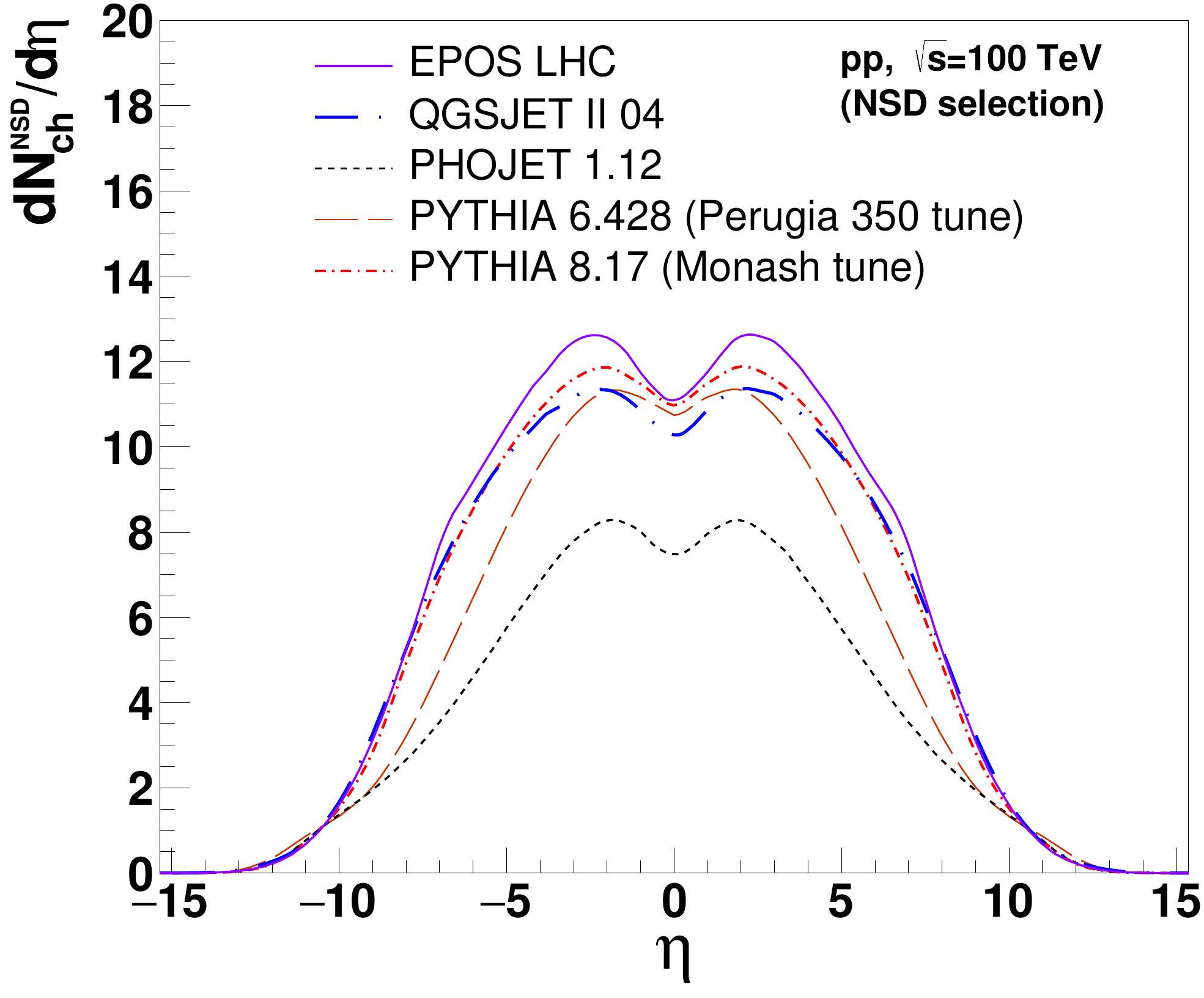}
\includegraphics[width=0.49\textwidth]{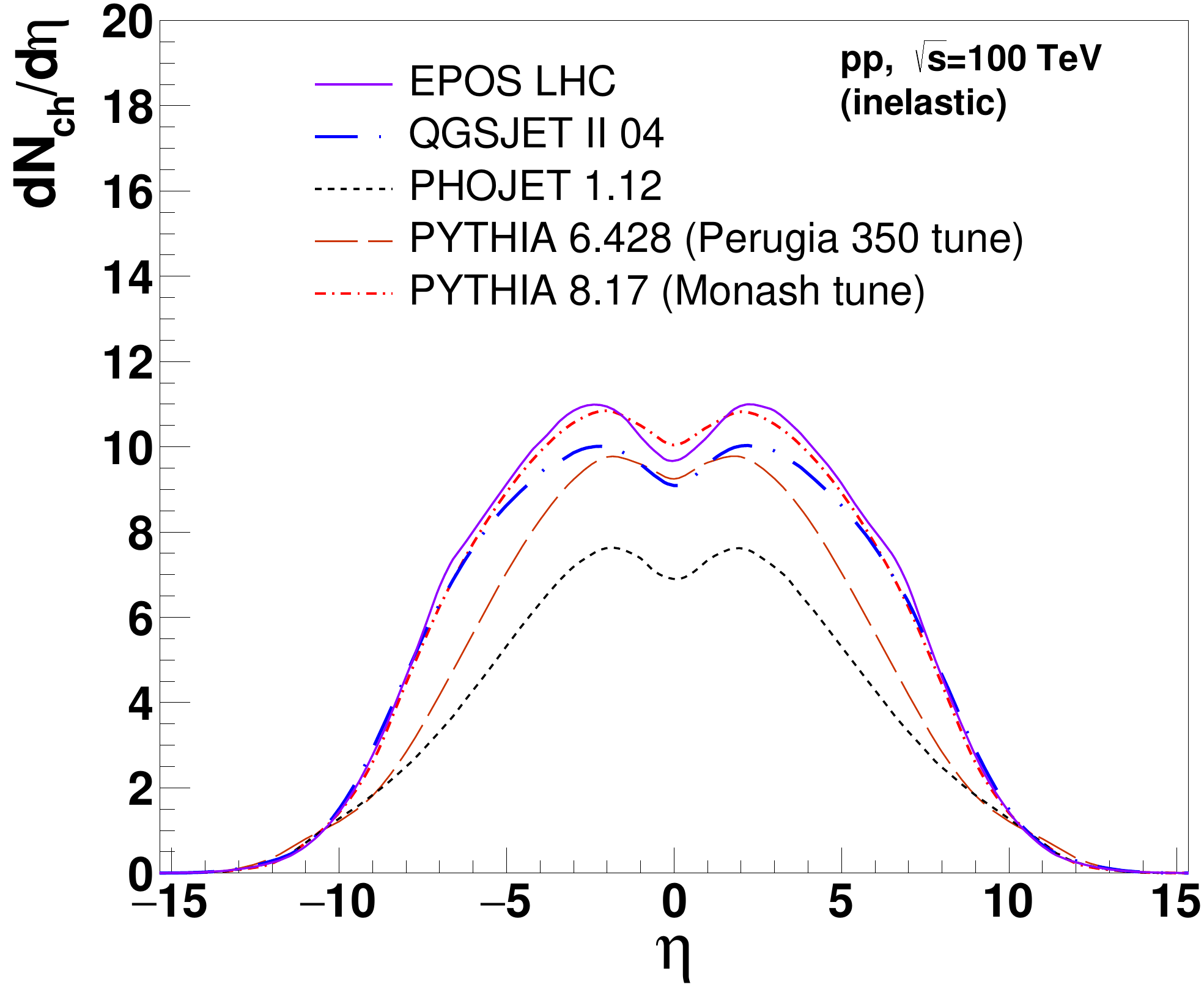}
\caption{Distributions of the pseudorapidity density of charged particles in non single-diffractive (left) and
  inelastic (right) \pp\ collisions at $\sqrts$~=~100~TeV, predicted by the different MCs considered in this work.} 
\label{fig:dNdeta_100TeV}
\end{figure}

Figure~\ref{fig:dNdeta_100TeV} shows the distribution of the number of charged particles produced in \pp\
collisions at 100~TeV per unit of pseudorapidity as a function of pseudorapidity ($\dNdeta$), predicted by the
different models in the range $|\eta|\lesssim 15$ (the beam {\it rapidity} at $\sqrts$~=~100~TeV is $y_{\rm beam} = \rm acosh(\sqrts/2.) \approx 11.5$).
The left plot shows the NSD distribution\footnote{In \pythia\ 6 and 8 this is achieved by directly
  switching off single-diffractive contributions via: \ttt{\scriptsize MSUB(92)=MSUB(93)=0}, and
  \ttt{\scriptsize SoftQCD:singleDiffraction=off}. For \phojet, \eposlhc\ and \qgsjetII\ only events MC-tagged as non-diffractive
or double diffractive are included.}, and the right one shows the inclusive inelastic
distribution which, including lower-multiplicity diffractive interactions, has a smaller average number of
particles produced. All models (except \phojet) predict about 10 charged particles at midrapidity ($\eta=0$). 
Taking an unweighted average of all the predictions (except \phojet\ which
is systematically lower by $\sim$40\%), we obtain:
$\dNdetaZeroNSD = 10.8 \pm 0.3$  and $\dNdetaZero = 9.6 \pm 0.2$.
The width of the central pseudorapidity ``plateau'' covers $\sim$10 units from $\eta
\approx -5$ to $\eta \approx +5$. At forward rapidities (equivalent to small $x\approx \pT/\sqrts\cdot
e^{-\eta}$) \pythia~6 and \phojet\ predict noticeably ``thinner'' distributions than the rest, due to lower underlying gluon
densities at $\pT \approx \rm Q_0$, than those from the NNPDF 2.3 LO set used in \pythia~8~\cite{Skands:2014pea}.
A significant fraction of the particles produced issue from the fragmentation of partons from semihard MPI, the
hardest partonic collision in the MB event producing only a small fraction of them. The fact that the \phojet\
particle yields are about $\sim$40\% lower than the rest of MCs is indicative of missing multiparton
contributions in this event generator.
\begin{figure}[htpb!]
\centering
\includegraphics[width=0.49\textwidth,height=7.cm]{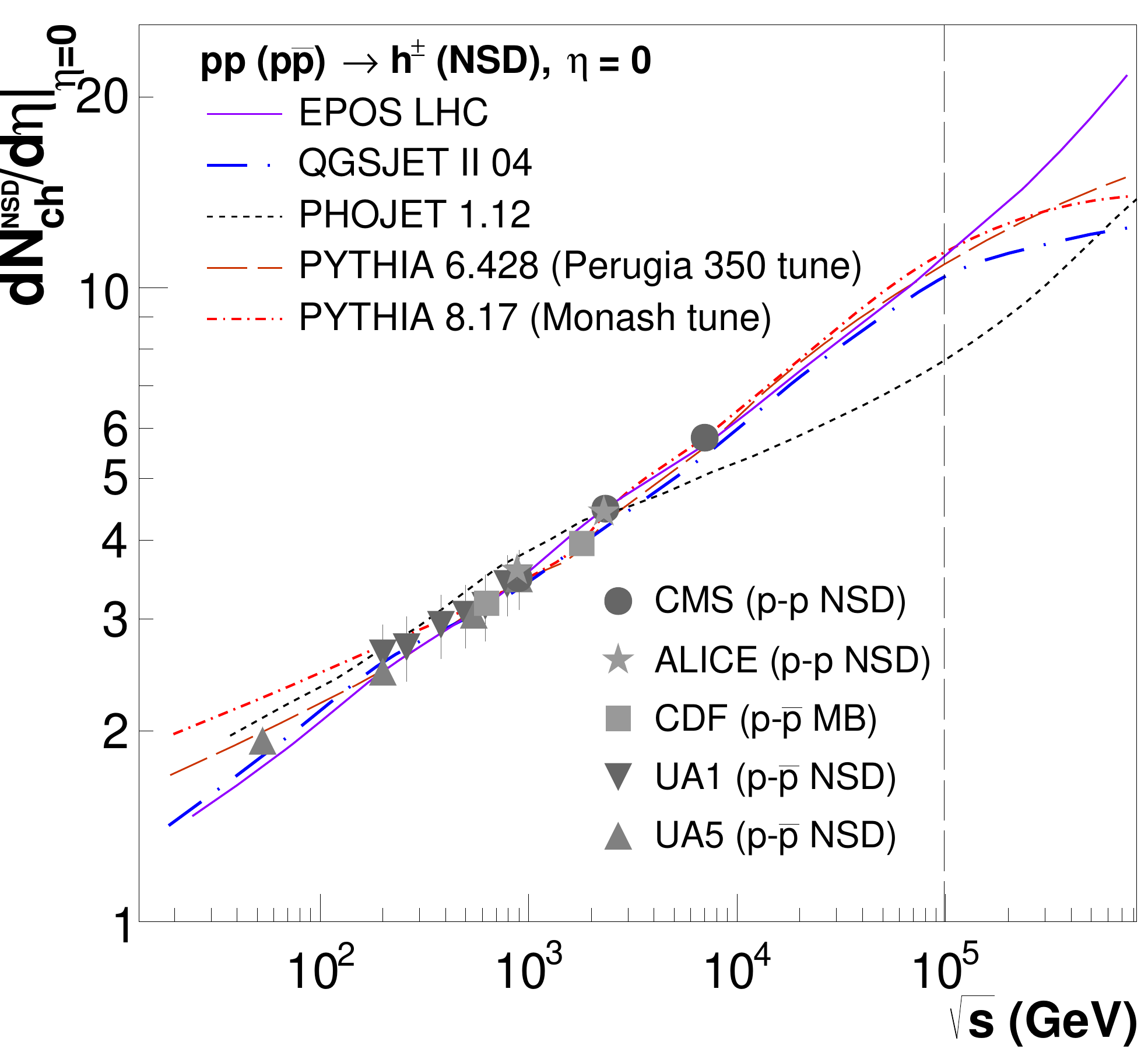}
\includegraphics[width=0.49\textwidth,height=7.cm]{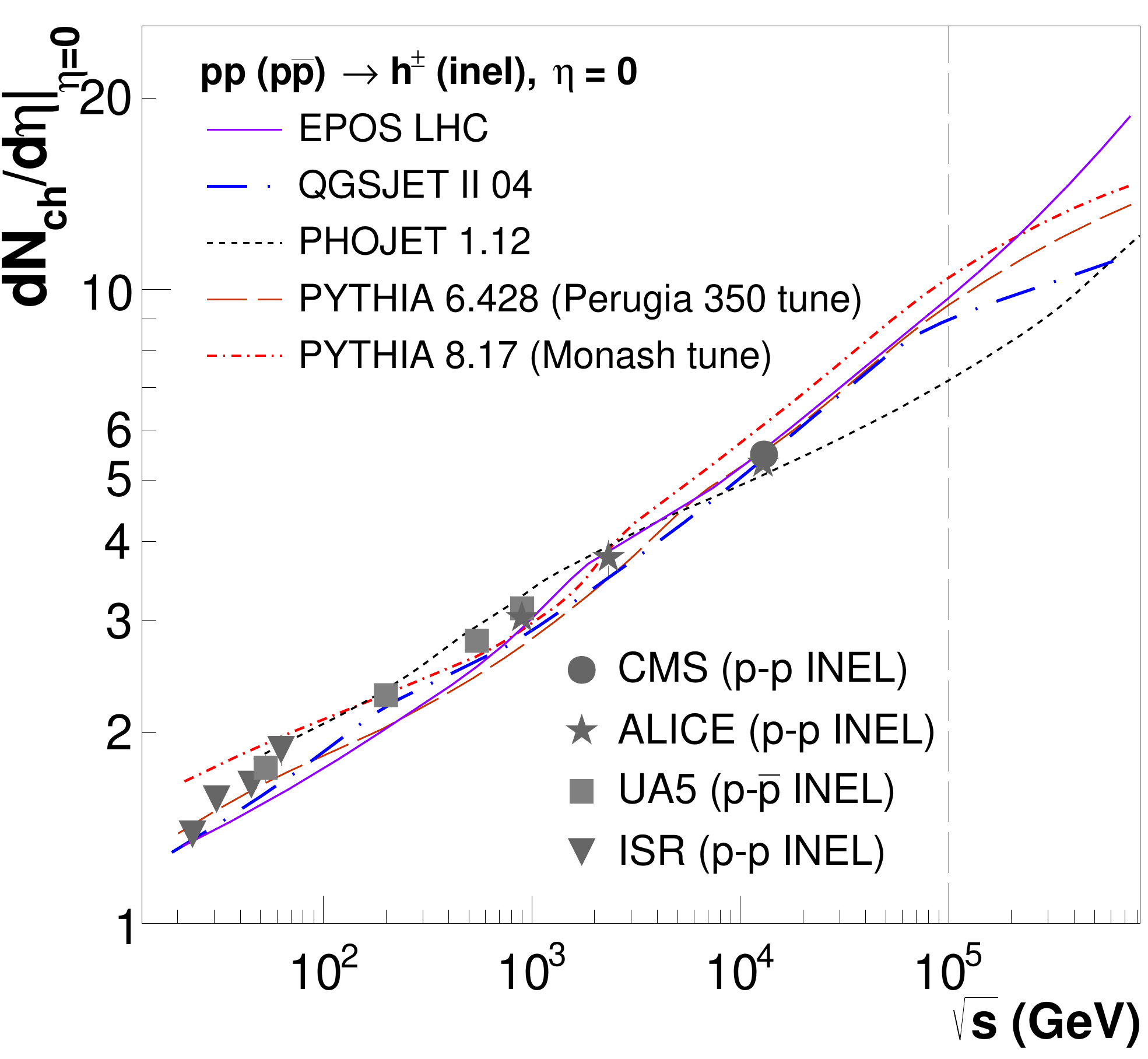}
\caption{Evolution of the charged particle pseudorapidity density at midrapidity, $\dNdetaZero$, as a
  function of collision energy, $\sqrts$, for non-single diffractive (left) and inelastic (right) \pp\
  collisions. The data points show existing collider 
  data~\cite{Albajar:1989an,Alner:1986xu,Aad:2010ac,Khachatryan:2010xs,Khachatryan:2010us,Khachatryan:2015jna}. 
  The vertical line indicates the FCC-hh/SppC energy at 100~TeV.}
\label{fig:dNdeta0_vs_sqrts}
\end{figure}
The \cm\ energy evolution of the charged hadron pseudorapidity density at $\eta = 0$ predicted by the different
models in the range $\sqrts$~=~10~GeV--800~TeV is presented in Fig.~\ref{fig:dNdeta0_vs_sqrts} compared
to the existing NSD (left panel) and inelastic (right panel) data measured at Sp$\bar{\mbox p}$S
(UA1~\cite{Albajar:1989an}, and UA5~\cite{Alner:1986xu}), Tevatron (CDF~\cite{Abe:1989td,Abe:1988yu}) and 
LHC (ALICE~\cite{Aamodt:2010ft,Aamodt:2010pp,Adam:2015pza}, ATLAS~\cite{Aad:2010ac} and
CMS~\cite{Khachatryan:2010xs,Khachatryan:2010us,Khachatryan:2015jna}) colliders. The expected increase in particle multiplicity
at midrapidity at 100~TeV is about a factor of two compared to the LHC results at 13~TeV
($\dNdetaZero$~=~5.31~$\pm$~0.18~\cite{Adam:2015pza},
5.49~$\pm$~0.17~\cite{Khachatryan:2015jna}). As aforementioned, the NSD selection has
central densities which are about 15\% larger than those obtained with the less-biased INEL trigger, which has less
particles produced on average as it includes (most of) diffractive production. All models (except \phojet,
whose results are not actually trustable beyond $\sqrts \approx$~75~TeV~\cite{Ralph}) more or less reproduce
the available experimental data up to LHC, and show a very similar trend with $\sqrts$ up to FCC-hh/SppC
energies. Beyond 100~TeV, however, \eposlhc\ tends to produce higher yields than the rest of MCs. 
It is worth to notice that, thanks to the LHC data, the differences among model predictions have been 
considerably reduced in comparison to the results of the pre-LHC models discussed in~\cite{d'Enterria:2011kw}.\\

The FCC-hh experiments aim at fully tracking coverage in the central
$|\eta|<5$ region. The total number of charged particles expected in the tracker system is obtained by
integrating the $\dNdeta$ distributions over that interval, which yields an average of
$\rm N_{_{\rm ch}}(\Delta\eta$=10$)\approx$~100. 
For the expected FCC-hh pileups, in the range $\cO{200-1000}$, this value implies that the trackers would sustain on
average a total number of 20--100 thousand tracks per bunch crossing. Such a value is of the same order of
magnitude as a {\it single} central Pb-Pb collision at LHC energies~\cite{Adam:2015ptt}, and thus perfectly
manageable for the high-granularity FCC-hh tracker designs.
Further integrating the $\dNdeta$ distributions over all pseudorapidities, one obtains the total number of charged
particles produced in an average \pp\ collision at 100~TeV. The \epos, \pythia~8 and \qgsjetII\ models
predict the largest total charged multiplicities, $\rm N_{_{\rm ch}}~(N_{_{\rm ch}}^{^{\rm NSD}})$~=~161~(184),
160~(170), 152~(172) respectively; followed by \pythia~6,  
$\rm N_{_{\rm ch}}~(N_{_{\rm ch}}^{^{\rm NSD}}) = 131~(150)$; and 
\phojet, $\rm N_{_{\rm ch}}~(N_{_{\rm ch}}^{^{\rm NSD}}) = 103~(111)$.

\subsection{Energy pseudorapidity density}

\begin{figure}[htpb!]
\centering
\includegraphics[width=0.49\textwidth]{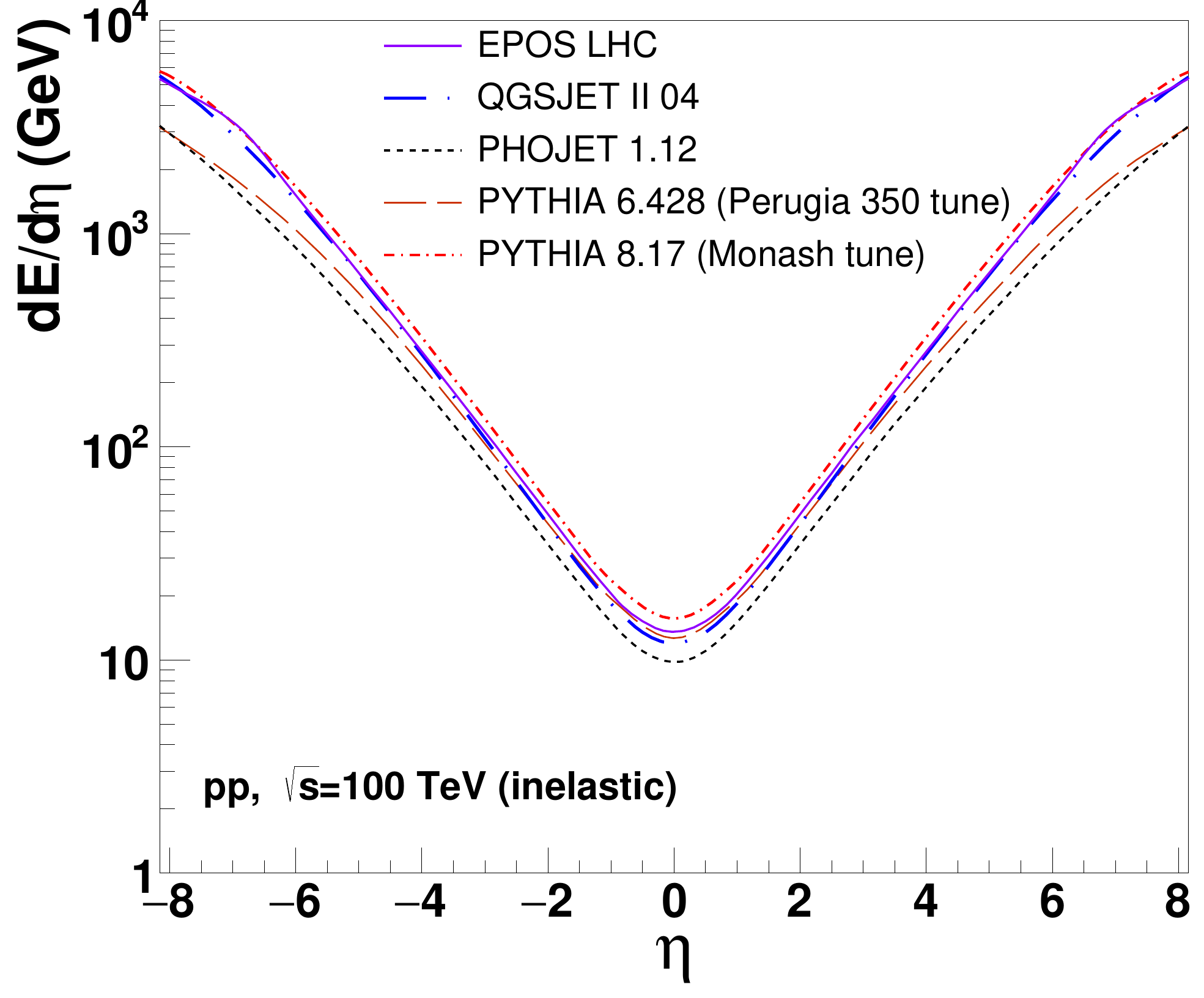}
\includegraphics[width=0.49\textwidth]{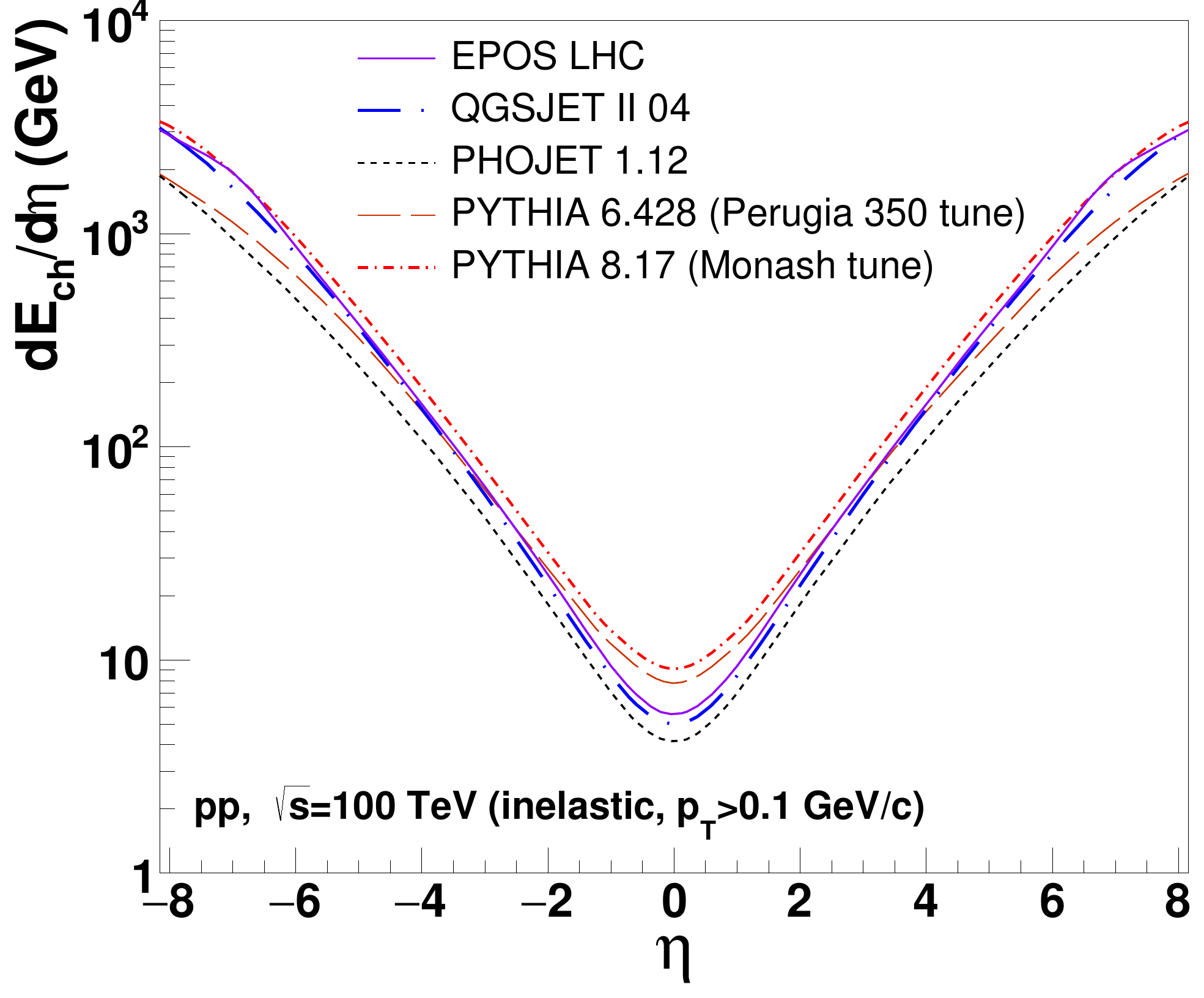}
\caption{Distribution of the energy pseudorapidity density of all particles (left) and of charged particles with
  $\pT>$~0.1~GeV/c (right) in inelastic \pp\ collisions at $\sqrts$~=~100~TeV, predicted by the different MCs considered in this work.}  
\label{fig:dEdeta_100TeV}
\end{figure}

Figure~\ref{fig:dEdeta_100TeV} shows the distributions of energy density as a function of pseudorapidity for the
total energy (left) and  for the energy carried by charged particles above a minimum $\pT = 100$~MeV/c
(right). \phojet\ predicts the lowest energy produced at all rapidities (consistent with the 
lower particle yields produced by the model), whereas \pythia~8 predicts the highest. At $\eta = 0$,
the total energy produced per unit rapidity is $\rm dE/d\eta = $~9.9, 12.2, 12.6, 13.7 and 15.6~GeV for
\phojet, \qgsjetII, \pythia~6, \eposlhc\ and \pythia~8 respectively. The same values at the forward edges of
typical detector coverages ($|\eta| = 5$) are  $\rm dE/d\eta\approx$~410, 525, 670, 700 and 760~GeV
for \phojet, \pythia~6, \qgsjetII, \eposlhc\ and \pythia~8 respectively. The trend for \pythia~6 is to
predict a smaller relative increase of energy density as a function of rapidity compared to the rest of
models due, again, to a more relatively depleted underlying gluon density at the increasingly lower $x$ values
probed at forward $\eta$.

\subsection{Multiplicity distribution}

\begin{figure}[htpb!]
\centering
\includegraphics[width=0.49\textwidth,height=7.cm]{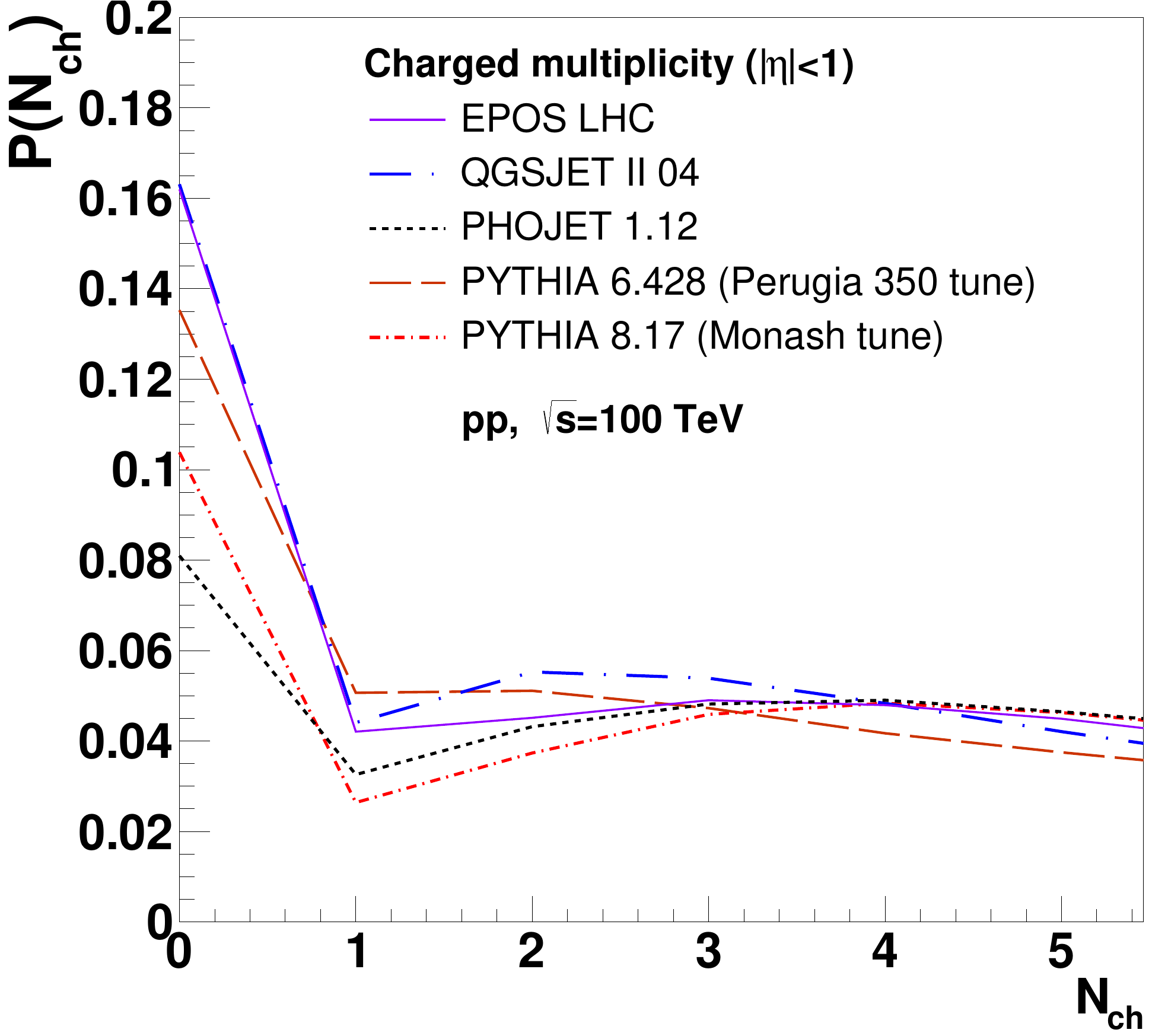}
\includegraphics[width=0.49\textwidth,height=7.cm]{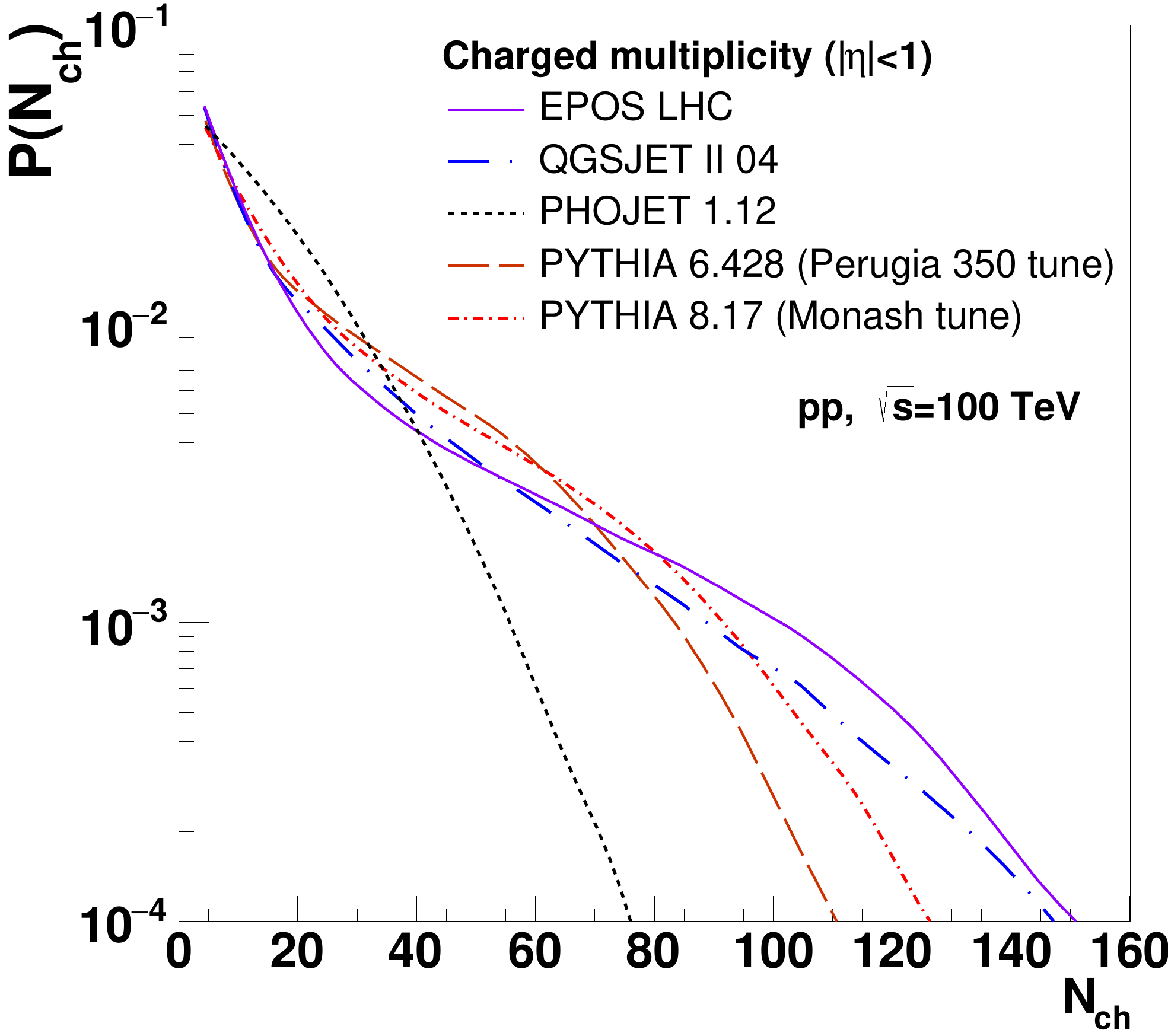}
\caption{Per-event charged particle probability (within $|\eta|<$~1) in inelastic 
\pp\ collisions at $\sqrts$~=~100~TeV: full distribution (right), zoom at low multiplicities $\PNch<5$ (left).}
\label{fig:PNch_100TeV}
\end{figure}

The multiplicity distribution $\PNch$, \ie\ the probability to produce $\rm N_{\rm ch}$
charged particles in a \pp\ event, provides important differential constraints on the internal details of the
hadronic interaction models. Figure~\ref{fig:PNch_100TeV} shows the distribution for charged particles
produced at central rapidities (within $|\eta|<1$) in inelastic \pp\ collisions at the FCC-hh/SppC. The  tail of the  
$\PNch$ distribution (right) gives information on the relative contribution of multiparton scatterings
(multi-Pomeron exchanges), whereas the low multiplicity part (left) is mostly sensitive to the contributions
from diffraction (single Pomeron exchanges). The various MCs considered predict quite different distributions
at both ends of the spectrum. The RFT-based models \eposlhc\ and \qgsjetII\ predict both higher yields at very low
($\rm N_{\rm ch}<3$) and very high ($\rm N_{\rm  ch}>100$) particle multiplicities, whereas \pythia~6 and 8 feature
higher yields in the intermediate region $\rm N_{\rm ch}\approx$~30--80. \phojet\ clearly produces too many
particles within $\rm N_{\rm  ch}\approx$~10--40, but much fewer at high multiplicities compared to the rest
of models (which is, again, indicative of missing MPI contributions in this MC generator).

\subsection{Transverse momentum distribution}

\begin{figure}[htpb!]
\centering
\includegraphics[width=0.46\textwidth,height=7.2cm]{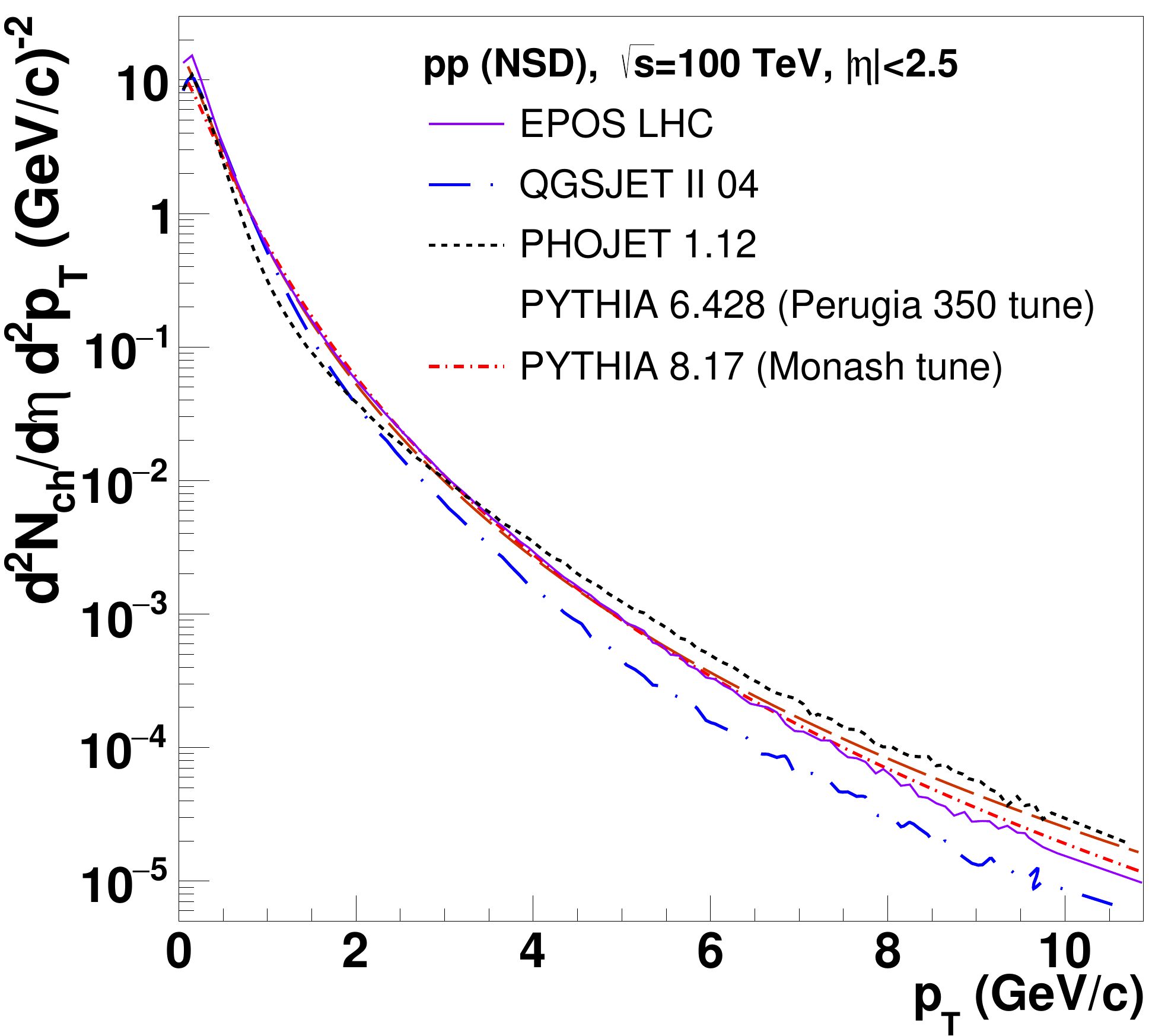} 
\includegraphics[width=0.53\textwidth,height=7.4cm]{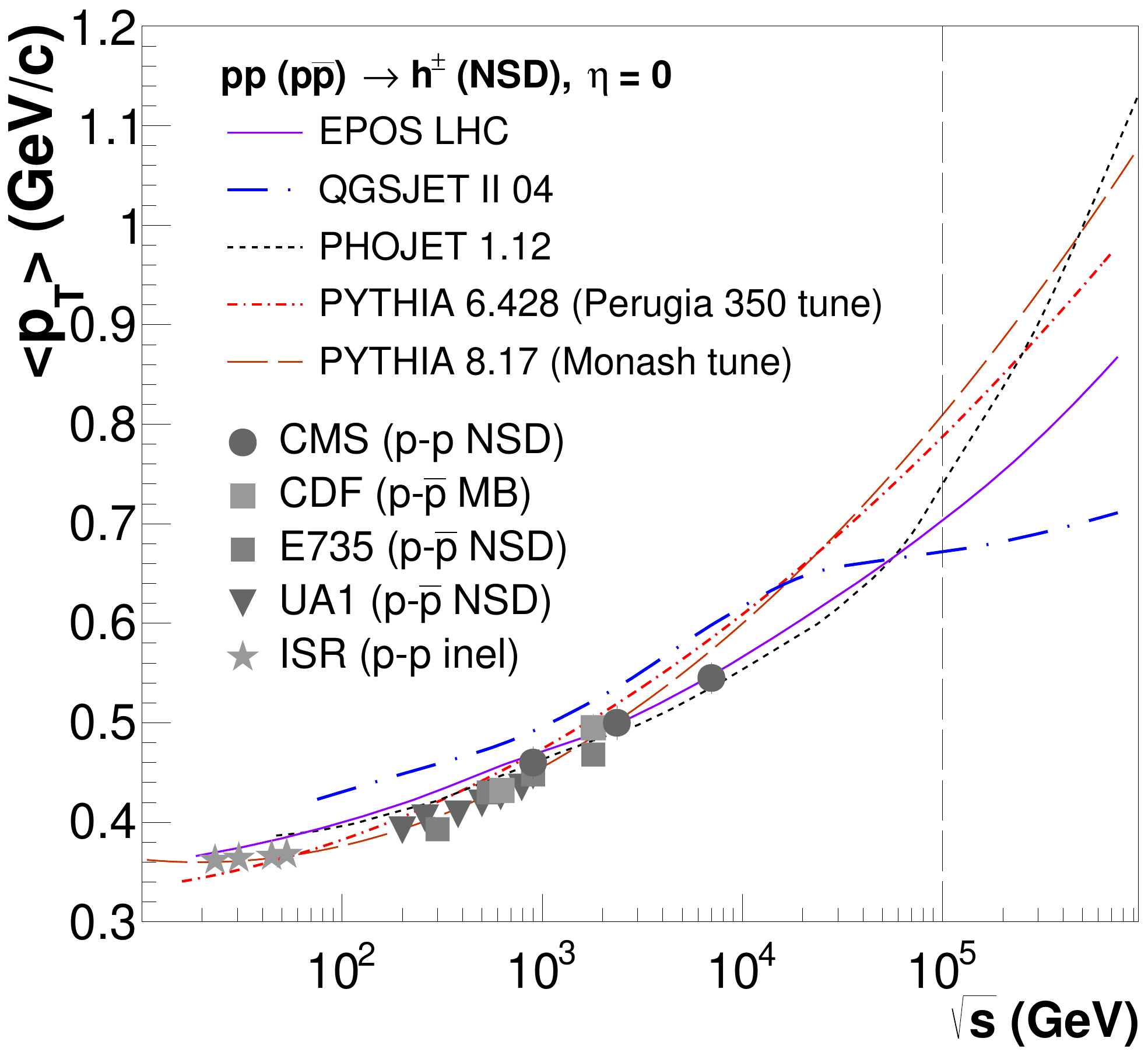}
\caption{Left: Transverse momentum spectrum in \pp\ collisions at
  $\sqrts$~=~100~TeV predicted by the different MCs considered in this work 
  (absolutely normalized at a common value at $\pT \approx 0.5$~GeV/c).
  Right: Evolution of 
  $\meanpt$ at midrapidity as a
  function of \cm\ energy $\sqrts$. Data points show existing collider
  results~\cite{Rossi:1974if,Albajar:1989an,Abe:1988yu,Khachatryan:2010xs,Khachatryan:2010us,Rossi:1974if,Alexopoulos:1988na},  
  and the vertical line indicates the FCC-hh/SppC energy at 100~TeV.} 
\label{fig:dNdpT}
\end{figure}

Figure~\ref{fig:dNdpT} (left) shows the $\pT$-differential distributions of charged particles at midrapidity
(within $|\eta|<2.5$) in \pp\ collisions at 100~TeV predicted by all models. 
All spectra have been absolutely normalized at their value at $\pT \approx 0.5$~GeV/c to be able to easily compare their 
shapes. Both \pythia~6 and 8 feature the largest yields at the high-$\pT$ end of the distributions (not shown here),
\qgsjetII\ features the ``softest'' spectrum, whereas \epos\ shows higher yields in the region
$\pT\approx$~1--5~GeV/c, due to collective partonic flow boosting the semihard region of the spectra, but then
progressively falls below the pure-pQCD \pythia\ MC generators. The \phojet\ spectrum has a more convex shape,
being comparatively depleted at intermediate $\pT\approx$~1--3~GeV/c but rising at its tail.
Studying the $\sqrts$-evolution of the average $\pT$ of the spectra provides useful (integrated) information.
At high energies, the peak of the perturbative cross section comes from interactions between partons whose 
transverse momentum is around the saturation scale, $\rm \meanpt \approx Q_{\rm sat}$, producing (mini)jets 
of a few GeV which fragment into lower-$\pT$ hadrons. As explained in the introduction, \pythia\ and \phojet\ MCs have an
energy-dependent $\pT$ cutoff that mimics the power-law evolution of $\rm Q_{\rm sat}$, while \epos\ and \qgsjet\ have
a fixed $\pT$ cutoff and low-$x$ saturation is implemented through corrections to the
multi-Pomeron dynamics. The different behaviours are seen in the $\sqrts$-evolution of the average
$\pT$ shown in Fig.~\ref{fig:dNdpT} (right). 
All MCs, but \qgsjetII, predict a (slow) power-law-like increase of $\meanpt$ with energy. Both \pythia~6 and 8
 {\textemdash} whose dynamics is fully dominated by (mini)jet production {\textemdash}  predict a higher $\meanpt$ than the rest of
models, yielding  $\meanpt\approx 0.82$~GeV/c at 100~TeV to be compared with $\meanpt$~=~0.73, 0.71 and 0.67~GeV/c
from \phojet, \eposlhc\ and \qgsjetII\ respectively. Above $\sqrts\approx$~20~TeV, \qgsjetII\ predicts a
flattening of $\meanpt$ whereas the \eposlhc\ evolution continues to rise due to final-state collective
flow which increases $\meanpt$ with increasing multiplicity.

\section{Summary}
\label{sec:summary}

In summary, the global properties of the final states produced in hadronic interactions of protons at
centre-of-mass energies of the of the CERN Future Circular Collider and of the IHEP Super proton-proton Collider,
have been studied with various Monte Carlo event
generators used in collider physics (\pythia~6, \pythia~8, and \phojet) and in ultrahigh-energy cosmic-rays
studies (\epos, and \qgsjet). Despite their different underlying modeling of hadronic interactions,
their predictions for proton-proton collisions at $\sqrts$~=~100~TeV are quite similar (excluding
\phojet, whose parameters have not been retuned with the collider data in the last 15 years).
Table~\ref{tab:MB_summary} lists  the basic kinematical observables predicted for \pp\ at 100~TeV by all MC
generators considered.\\

\begin{table}[htbp]
\centering
\begin{tabular}{l|c|c|c|c|c|c}\hline\hline
 &\pythia~6& \pythia~8 & \eposlhc & \qgsjet~II\ & \phojet &  \hspace{0.6cm}Average$^\star$\hspace{0.6cm} \\ \hline
$\sigmainel$ (mb)  & $106.9$ & $107.1$ & $105.4$ & $104.8$ & $103.1$ & $105.1 \pm 2.0$\\ \hline
$\rm N_{_{\rm ch}} (N_{_{\rm ch}}^{^{\rm NSD}})$ &  131 (150) & 160~(170) & 161 (184) & 152 (172) &  101 (121) & 150 (170) $\pm$ 20\\
$\dNdetaZero$      & $ 9.20 \pm 0.01$ & $10.10 \pm 0.06$ & $ 9.70 \pm 0.16$ & $9.10 \pm 0.15$ & $6.90 \pm 0.13$ & $9.6 \pm 0.2$ \\
$\dNdetaZeroNSD$   & \hspace{0.2cm} $10.70 \pm 0.06$ \hspace{0.2cm} & \hspace{0.2cm} $10.90 \pm 0.06$
\hspace{0.2cm} & \hspace{0.2cm} $11.10 \pm 0.18$ \hspace{0.2cm} & \hspace{0.2cm} $10.30 \pm 0.17$ \hspace{0.2cm} & \hspace{0.2cm} $7.50 \pm 0.15$ \hspace{0.2cm} & \hspace{0.2cm} $10.8 \pm 0.3$ \hspace{0.2cm}\\ \hline
$\rm dE/d\eta|_{\eta=0}$ (GeV) & $12.65 \pm 0.07$ & $ 15.65 \pm 0.02$  & $ 13.70 \pm 0.02$ & $ 12.2 \pm 0.02$ & $9.9 \pm 0.01$ & $ 13.6 \pm 1.5$ \\
$\rm dE/d\eta|_{\eta=5}$ (GeV) & $525 \pm 4$ & $760 \pm 1$ & $ 700 \pm 1$ & $ 670 \pm 1$ & $410 \pm 1$ & $ 670 \pm 70$ \\ \hline
$\rm P(N_{ch}< 5)$  & $0.28$ & $0.22$ & $0.35$ & $0.36$ & $0.25$ & 0.30 $\pm$ 0.03 \\
$\rm P(N_{ch}>100)$ & $3.3\cdot10^{-3}$ & $0.011$ & $0.025$ & $0.018$ & $10^{-5}$ & 0.015 $\pm$ 0.05 \\ \hline
$\meanpt$ (GeV/c)   & $0.80 \pm 0.02$ & $0.84 \pm 0.02$ & $0.71 \pm 0.02$ & $0.67 \pm 0.02$ & $0.73 \pm 0.02$ & $0.76 \pm 0.07$ \\\hline\hline
\end{tabular}
\caption{Comparison of the basic properties of particle production in \pp\ collisions at $\sqrts$~=~100~TeV,
  predicted by \pythia~6 and 8, \eposlhc, \qgsjetII, and \phojet: Inelastic cross section $\sigmainel$;
  total charged multiplicities ($\rm N_{_{\rm ch}}$), and pseudorapidity charged particle densities at
  midrapidity ($\dNdetaZero$) for inelastic and NSD selections; energy densities at midrapidity ($\rm
  dE/d\eta|_{\eta=0}$), and at more forward rapidities ($\rm dE/d\eta|_{\eta=5}$); typical values of the charged
  multiplicity probabilities $\rm P(N_{ch})$ (over $|\eta|<1$) for low and high values of $\rm N_{ch}$; and
  mean charged particle transverse  momentum $\meanpt$ over $|\eta|<2.5$. The quoted uncertainties on the individual
  predictions are just the MC statistical ones. The last column indicates the average of all MCs
  (except \phojet)$^\star$\ for each observable, with uncertainties approximately covering the range of the predictions.} 
\label{tab:MB_summary}
\end{table}

The averages of all MC predictions (except \phojet) for  the different observables are:
(i) \pp\ inelastic cross sections $\sigmainel$~=~105~$\pm$~2~mb (to be compared with
$\sigmainel \approx$~72~mb at the LHC(13~TeV), \ie\ a $\sim$45\% increase),
(ii) total charged multiplicity $\rm N_{_{\rm ch}}~(N_{_{\rm ch}}^{^{\rm NSD}})$~=~150~(170)~$\pm$~20,
(iii) charged particle pseudorapidity density at midrapidity $\dNdetaZero = 9.6 \pm 0.2$ 
(to be compared with the LHC(13~TeV) result of $\dNdetaZero$~=~5.4~$\pm$~0.2, \ie\ an increase of
$\sim$80\%), and $\dNdetaZeroNSD = 10.8 \pm 0.3$ for the NSD selection, 
(iv) energy density at midrapidity $\rm dE/d\eta|_{\eta=0} =  13.6 \pm 1.5$~GeV,
and energy density at the edge of the central region $\rm dE/d\eta|_{\eta=5} = 670 \pm 70$~GeV, and 
(v) average transverse momenta at midrapidities $\meanpt = 0.76 \pm 0.07$~GeV/c (to be compared with $\meanpt$~=~0.55~$\pm$~0.16
at the LHC(8~TeV), \ie\ a $\sim$40\% increase).
The per-event multiplicity probabilities $\PNch$, have been also compared: \eposlhc\ and
\qgsjetII\ both predict higher yields at very low ($\rm N_{\rm ch}<3$) and very high ($\rm N_{\rm  ch}>100$)
particle multiplicities, whereas \pythia~6 and 8 feature higher yields in the intermediate region $\rm N_{\rm
  ch}\approx$~30--80. These results are useful to estimate the expected detector occupancies and energy
deposits from pileup collisions at high luminosities of relevance for planned FCC-hh/SppC detector designs.\\

\paragraph*{\bf Acknowledgments  {\textemdash} }
We are grateful to Peter Skands for useful discussions and feedback to a previous version of this document,
and to Ralph Engel for valuable discussions.


\end{document}